# Development of pediatric myeloid leukemia may be related to the repeated bone-remodeling during bone-growth

Jicun Wang-Michelitsch[1]*, Thomas M Michelitsch[2]

[1] Independent researcher

[2] Sorbonne Université, Institut Jean le Rond d'Alembert, CNRS UMR 7190 Paris, France

## Abstract

Acute myeloid leukemia (AML) and chronic myeloid leukemia (CML) are two major forms of leukemia developed from myeloid cells (MCs). To understand why AML and CML occur in children, we analyzed the causes and the mechanism of cell transformation of a MC. **I.** DNA changes in cells are generated as consequences of cell injuries and DNA injuries. For the MCs and hematopoietic stem cells (HSCs) in marrow cavity, repeated bone-remodeling during bone-growth and bone-repair may be a source of cell injuries. **II.** As a type of blood cell, a MC may have higher survivability from DNA changes and require obtaining fewer cancerous properties for cell transformation than a tissue cell. **III.** Point DNA mutations (PDMs) and chromosome changes (CCs) are the two major types of DNA changes. CCs have three subtypes by their effects on a cell: great effect CCs (GECCs), mild-effect CCs (MECCs), and intermediate-effect CCs (IECCs). A GECC affects one or more genes and can alone trigger cell transformation. PDMs/MECCs are mostly mild and can accumulate in cells. Some of the PDMs/MECCs contribute to cell transformation. An IECC affects one or more genes and participates in cell transformation. **IV.** Based on **II** and **III**, we hypothesize that a MC may have two pathways on transformation: a slow and an accelerated. Slow pathway is driven by accumulation of PDMs/MECCs through many generations of cells. Accelerated pathway is driven by accumulation of PDMs/MECCs/IECC(s) through a few generations of cells. In both pathways, long-term accumulation of DNA changes occurs in regenerable HSCs. A transformation via slow pathway occurs at old age; whereas that via accelerated pathway occurs at any age. Thus, CML and pediatric AML may develop via accelerated pathway, and adult AML may develop via slow or accelerated pathway. **In conclusion**, pediatric AML and CML may develop as a result of cell transformation of a MC via accelerated pathway; and the repeated bone-remodeling during bone-growth may be a trigger for the cell transformation of a MC in a child.

**This paper has the following structure:**

**I.    Introduction**

**II.   Acute myeloid leukemia (AML) and chronic myeloid leukemia (CML) affect both adults and children**

  2.1 AML occurs at all ages but mostly in adults
      2.1.1    Different subtypes of AML occur at different ages
      2.1.2    Different subtypes of AML have different forms of recurrent DNA changes
  2.2 CML occurs in both children and adults
      2.2.1    Ph translocation is an essential DNA change in CML development
      2.2.2    CML is a form of myeloid proliferative neoplasm (MPN)
  2.3 Myelodysplastic syndrome (MDS) occurs mainly in old people
  2.4 Juvenile myelomonocytic leukemia (JMML) is a type of MDS/MPN

**III.  Development of myeloid cells (MCs) in bone marrow**

  3.1   Ancestors of MCs: hematopoietic stem cells (HSCs)
  3.2   Developing MCs: including progenitor cells, blast cells, and pro-cytes
  3.3   Mature MCs: including granulocytes, monocytes, red cells, and megakaryocytes (platelets)

**IV.   A potential source of injuries of hematopoietic cells (HCs) in marrow: the repeated bone-remodeling during bone-growth and bone-repair**

**V.    Generation and accumulation of DNA changes in HCs is a result of repeated cell injuries and repeated cell proliferation**

**VI.   Different types of DNA changes have different effects on a cell**

  6.1  Effect of a point DNA mutation (PDM) on a cell
  6.2  Three types of chromosome changes (CCs) by their effects on a cell
      6.2.1    Great-effect CCs (GECCs)
      6.2.2    Mild-effect CCs (MECCs)
      6.2.3    Intermediate-effect CCs (IECCs)

**VII.  A MC may have higher survivability from DNA changes than a tissue cell**

  7.1 Higher tolerance of a MC to DNA changes than a tissue cell
  7.2 A developing MC may have higher survivability from DNA changes than a mature MC

**VIII. A MC may require obtaining fewer cancerous properties for transformation than a tissue cell**

**IX.   Our hypothesis: a MC may have two pathways on transformation: a slow and an accelerated**

  9.1 Slow pathway: by accumulation of PDMs and MECCs through many generations of cells
  9.2 Accelerated pathway: by accumulation of PDMs, MECCs, and IECC(s) through a few generations of cells

## I. Introduction

Acute myeloid leukemia (AML) and chronic myeloid leukemia (CML) are two major forms of leukemia arising from myeloid cells (MCs) in marrow. "MCs" include all the cells of myeloid linage at any developing stages. AML and CML occur mostly in adults but also in children. Myelodysplastic syndrome (MDS) is a type of pre-leukemia that occurs mainly in old people. In updated WHO classification of AML and MDS in 2016, some subtypes of AML are renamed by their specific DNA changes, for example, "AML with t(6;9)" and "AML with t(8;21)" (Arber, 2016). The novel classification of AML is important in clinic for making a more precise diagnosis and prognosis of an AML. Nevertheless, in this paper, we discuss AML by using the FAB classification of AML subtypes (M0-M7). The classic name of an AML subtype gives information of the pathology and the cell of origin of an AML. On the causing factors for leukemia, we have recently proposed a hypothesis: repeated bone-remodeling during bone-growth may be a source of cell injuries of the hematopoietic cells (HCs) in marrow (Wang-Michelitsch, 2018a). HCs include hematopoietic stem cells (HSCs), MCs, and the lymphoid cells (LCs) in marrow. We discuss in this paper the developing mechanisms of AML and CML by this hypothesis. Our discussion is driven by the following questions:

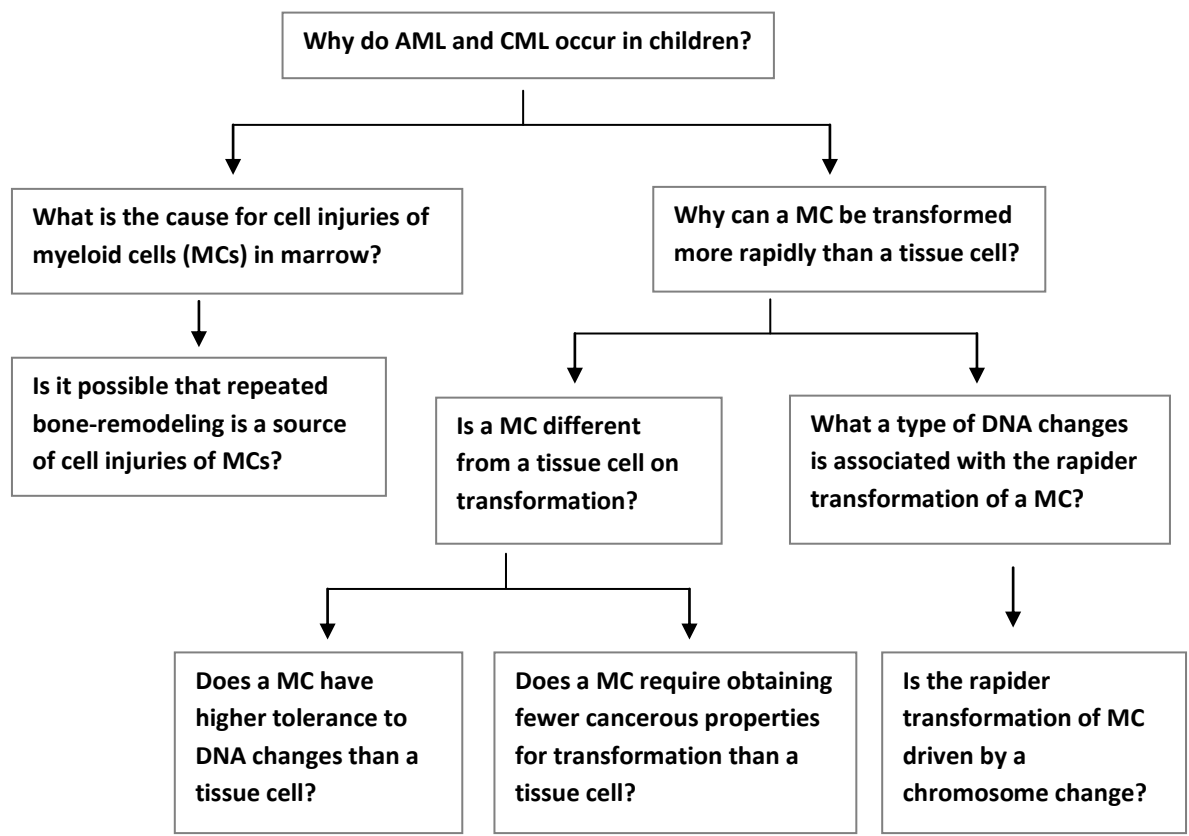

We aim to show by our discussion that: pediatric AML and CML may develop as a result of cell transformation of a MC via accelerated pathway; and the repeated bone-remodeling during bone-growth may be related to the cell transformation of a MC in a child. We use the following abbreviations in this paper:

AML: acute myeloid leukemia
ALL: acute lymphoblastic leukemia
BL: Burkitt lymphoma
CML: chronic myeloid leukemia
GECC: great-effect chromosome change
HC: hematopoietic cell
HSC: hematopoietic stem cells
IECC: intermediate-effect chromosome change
JMML: Juvenile myelomonocytic leukemia
LC: lymphoid cell
LCCI: loss of cell-contact inhibition
LPC: lymphoid progenitor cell

CLL: chronic lymphocytic leukemia
DLBCL: diffuse large B-cell lymphoma
DSB: double-strand break of DNA
GMP: granulocyte-monocyte progenitor
MC: myeloid cell
MDS: myelodysplastic syndrome
MECC: mild-effect chromosomal change
MEP: megakaryocyte-erythrocyte progenitor
MPC: myeloid progenitor cell
MPN: myeloid proliferative neoplasm
PDM: point DNA mutation
SIM: stimulator-independent mitosis

## II. Acute myeloid leukemia (AML) and chronic myeloid leukemia (CML) affect both adults and children

AML and CML affect both adults and children, but MDS affects mainly the people older than age 60 (Figure 1). AML can occur in infants, and has a small peak before age 2. 10% of CML patients are children and adolescents; however CML occurs rarely in children under age 4. Pediatric AMLs can be different from adult cases by disease presentation, pathology, and prognosis. Leukemia patients at different ages need to be given different treatments, thus they are regrouped often by age: infants (age 0-2), children (age 3-14), adolescents (age 15-19), young adults (age 20-39), and adults (age > 39). Pediatric patients are the patients that are younger than age 15 (Kent, 2009). AML and MDS have much higher incidences than CML. In USA, the incidences of AML, CML and MDS in 2013 are respectively 12950, 5980, and 14100. All these three forms of leukemia have higher incidences in males than in females.

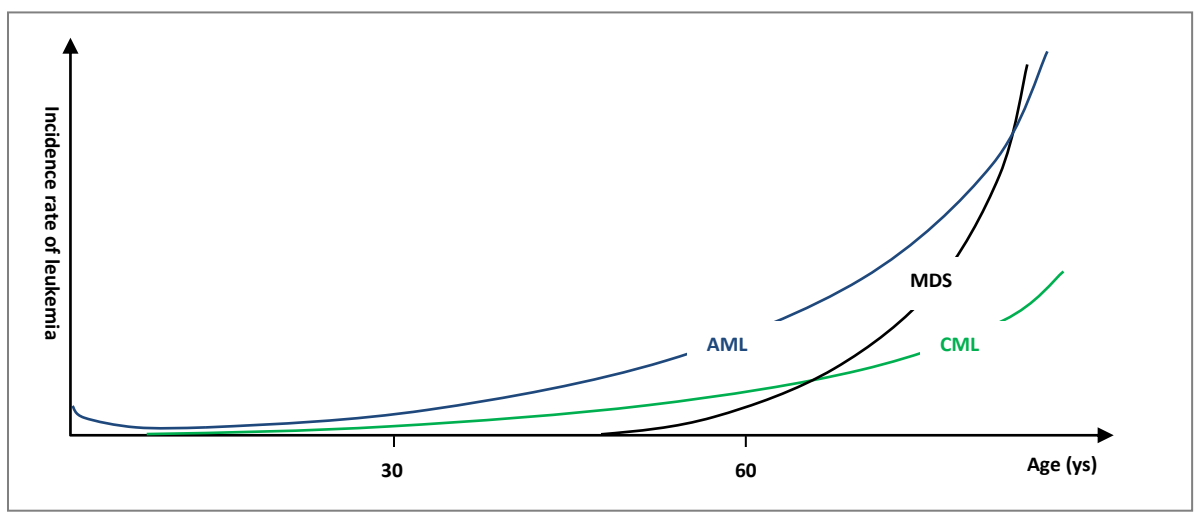

Figure 1. Age-specific incidences of AML, CML and MDS (a schematic graph)

AML (**blue line**) and CML (**green line**) occur mostly in adults but also in children. AML can occur in infants, and has a small peak before age 2. CML occurs rarely before age 4. MDS occurs mainly in old people (**black line**). AML and MDS have much higher incidences than CML.

### 2.1 AML occurs at all ages but mostly in adults

AML is a form of blood cancer that exhibits as abnormal cloning and differentiation of myeloid blast cells in marrow. AMLs account for 80% of adult leukemia cases and 15% of pediatric cases. AML has increasing incidence with age (Figure 1). However, AML has a small peak before age 2 (Pérez-Saldivar, 2011). It is not clear yet what is the main causing factor for AML. Possible causes include genetic disorders and exposures to ionizing radiation or organic solvents. For infant AML (IAML), exposures to radiation/chemicals or smoking of the mother during pregnancy may be a cause in some cases. The infants with some hereditary conditions, such as Down syndrome, Fanconi anemia, Diamond-Blackfan syndrome, and Li-Fraumeni syndrome, have high risk of IAML development (Saida, 2017).

#### *2.1.1 Different subtypes of AML occur at different ages*

By pathology, AML is classified into **M0-M7** eight subtypes. Diagnosis of a subtype of AML is based on the characteristics of leukemic cells on morphology, immunophenotype, and cytogentic changes (Saultz, 2016). The names of AML subtypes are listed here: **M0:** Undifferentiated acute myeloblastic leukemia; **M1**: Acute myeloblastic leukemia with minimal maturation; **M2**: Acute myeloblastic leukemia with maturation; **M3**: Acute promyelocytic leukemia (APL); **M4**: Acute myelomonocytic leukemia; **M4eos**: Acute myelomonocytic leukemia with eosinophilia; **M5**: Acute monocytic leukemia; **M6**: Acute erythroid leukemia; and **M7**: Acute megakaryoblastic leukemia. Different subtypes of AML have different incidence rates in population and have different age-distributions. **M0**, **M1** and **M2** are all myeloblastic leukemia and together represent 45% of AMLs. The incidence rates of **M3, M4/M4eos,** and **M5** are respectively 10%, 25%, and 10%. **M6** and **M7** are rare (Santos, 2009; Masetti, 2015).

**M0** occurs mostly in infants and older adults (Venditti, 1997). **M1** and **M2** are the most common subtypes of AML in children and adults, but they are rare in infants (Table 1). **M3** occurs mostly at age 35-40. 50% of IAMLs are **M4** and **M5**, and 20% are **M7** (Masetti, 2015). **M6** occurs mostly in people older than age 60 (Santos, 2009). **M7** is the most common subtype of AML in Down's children. A child with Down syndrome may have 16-fold higher risk than other children on developing AML. Familiar AMLs occur mostly before age 3. In general, pediatric AMLs have better prognosis than adult cases after treatment. However, most IAMLs are chemotherapy-resistant. The 5-year overall survival rates (OS) in infants, children, and adults (age >15) are respectively 10%-30%, 50%, and 30%-40% (Master, 2016). Among all subtypes of AML, APL (**M3**) has the best prognosis (Kansal, 2016).

#### *2.1.2 Different subtypes of AML have different forms of recurrent DNA changes*

More than 300 forms of chromosome changes (CCs) and molecular mutations have been identified in AML cells. CCs have higher occurrences in pediatric AMLs: in 70%-85% of pediatric AMLs whereas in 50% of adult cases (Kumar, 2011). Different subtypes of AML often have different recurrent forms of CCs in tumor cells (Kansal, 2016). The most frequent forms of CCs in each subtype are respectively: **M0**: complex karyotypes; **M1**: t(9;22), (+8), (-5), (-5q), and abn(3q); **M2**: t(8; 21), t(3;21), and t(6;9); **M3**: t(15;17) and t(11;17); **M4** and **M4eo**: inv (16); **M5**: t(4;11), t(9; 11), and t(11;19); **M7**: t(1;22) and (+21) (Table 1). Some forms of CCs seem to be age-specific. For example, t(1;22), t(5;11), t(7;12), and complex karyotypes are forms of CCs mainly seen in infants. 5%-30% of IAMLs have t(1;22) and 14% have complex karyotypes. t(1;22) is seen only in new babies whereas complex karyotypes are observed in infants older than 6 months. MLL rearrangements are forms of CCs often seen in IAMLs (Balgobind, 2011). They include t(4;11) (MLL/AF4), t(9;11) (MLL/AF9), and t(11;19). 35%-50% of IAMLs have t(4;11). Differently, t(8;21) is found mainly in older children and adults. Forms such as (-5), (-5q), and abn (3q) are seen only in adults.

Gene mutations are also common in AMLs, and the most frequent ones are mutations on *NPM1* (in 40%-60% of AMLs), *FLT-ITD* (30%), and *CEBPA* (15%) (Table 1) (Kumar, 2011; Nardi, 2016). Familiar AMLs have often mutations on *RUNX1*, *CEBPA,* and *MLL* (Stieglitz, 2013). Among all these forms of CCs and gene mutations, some are related to disease prognosis. The forms associated with good prognosis of AML include t(15;17), t(11;17), t(8;21) and t(3;21), inv(16), and *CEBPA* mutations (Kumar, 2011). Those associated with intermediate prognosis of AML include t(9; 11), (+8), (-y), and mutations of *NPM1* and *FLT-ITD*. Those associated with poor prognosis are t(6;9), t(9;22), (-5), (-7), inv (3) (GATA2 insufficiency), abn(3q), complex karyotypes, and mutations of *MLL, C-kit, TP53*, and *RUNX1* (Foucar, 2015).

**Table 1. Different subtypes of AML have different recurrent DNA changes**

|  | Infants (< age 2) | Children (age 3-14) | Adults (> age 15) |
| --- | --- | --- | --- |
| **Main subtypes of AML** | **M0, M4/M4eo, M5,** and **M7** | **M1, M2, M4,** and **M5** | **M1, M2, M3,** and **M6** |
| **Chromosome changes** | **M0**: complex karyotypes<br>**M4**: inv (16)<br>**M5**: t(4;11), t(9;11), and t(11;19)<br>**M7**: t(1;22) and (+21) | **M1**: t(9;22) and ( +8)<br>**M2**: t(8;21), t(3;21), and t(6;9)<br>**M4**: inv (16)<br>**M5**: t(4;11), t(9;11), and t(11;19) | **M1**: t(9;22), (+8), (-5), (-5q), and abn(3q)<br>**M2**: t(8;21), t(3;21), and t(6;9)<br>**M3**: t(15;17) and t(11;17) |
| **Gene mutations** | *RUNX1* and *MLL* in familiar AML | *NPM1,FLT-ITD,*and *CEBPA* | *NPM1, FLT-ITD,* and *CEBPA* |
| **5-year survival rates** | 10%-30% | 50% | 30%-40% |

## 2.2 CML occurs in both children and adults

CML is an indolent form of blood cancer and it exhibits as over-production of all types of blood cells in myeloid lineage (Sarwar, 2015). CML occurs in both children and adults (Figure 1). 10% of CML patients are children and adolescents (Höglund, 2015). However, CML occurs rarely in young children (< age 4) (Homans, 1984). About 30% of CML patients develop AML after 3-4 years of CML. Progression of CML into AML has three phases: chronic phase (CML-CP), accelerating phase (CML-AP), and transforming phase (CML-TP, called also blast crisis). 85% of CMLs are diagnosed at chronic phase. However, pediatric CMLs are often diagnosed at accelerating phase or transforming phase. Pediatric CML seems to be more aggressive than adult CML.

### 2.2.1 Ph translocation is an essential DNA change in CML development

Philadelphia (Ph) translocation (t(9;22)) is a characteristic DNA change in CML. 95% of CMLs have Ph translocation in leukemia cells. Ph translocation results in generation of fusion gene of *BCR-ABL1*. ABL1 is a tyrosine kinase that mediates the signal transduction of some stimulators that can activate cell proliferation. Fusion of *ABL1* gene with *BCR* gene results in permanent activation of ABL1 tyrosine kinase and unlimited cell proliferation of MCs (Bennour, 2015). Thus, inhibiting the activity of tyrosine kinase by specific inhibitors is a way to slow down the cell proliferation of MCs in CML. Clinical studies have shown that tyrosine kinase inhibitors (TKIs) including imatinib are effective drugs on controlling CML. CML-CP is curable by TKIs (Jain, 2016).

However, Ph translocation is not specific for CML. Ph translocation is also found in 10%-20% of cases of acute lymphoblastic leukemia (ALL). In these ALLs, Ph translocation contributes to the cell transformation of lymphoblast. Ph translocation is the main form of CC in CML at chronic phase. However, secondary DNA changes can be generated during the progression of CML. Some of the secondary DNA changes may lead to blast crisis of CML. These secondary DNA changes include t(8;21), second Ph translocation, isochromosome 17q, and (+19) (Bozkurt, 2013).

### 2.2.2 CML is a form of myeloid proliferative neoplasm (MPN)

CML is a proliferative disease, as a type of MPN. Other types of MPN include polycythemia vera (PV), essential thrombocytosis (ET), primary myelofibrosis (PMF), chronic eosinophilic leukemia (CEL), and chronic neutrophilic leukemia (CNL) (Clara, 2016). Each type of MPN is a result of cloning proliferation of a MC. CML occurs in both children and adults, but other types of MPN develop mainly in old people. In addition, CML affects all the MCs in five myeloid lineages, but each of PV, ET, CEL, and CNL affects only one of the five lineages of MCs. These types of MPN do not have Ph translocation in leukemia cells. 95% of PV/ET and 50% of PMF have *JAK2* V67F mutations or exon 12 mutations. These mutations lead to permanent activation of JAK 2 kinase in cells (Bose, 2017). JAK2 inhibitors such as ruxolitinib have been developed for treating MPNs.

## 2. 3 Myelodysplastic syndrome (MDS) occurs mainly in old people

MDS is a group of neoplasms characterized by peripheral cytopenia as a result of ineffective hematopoiesis in marrow. MDS is a type of pre-leukemia because MDS becomes AML when the blast count in marrow increases to 20%. In updated WHO classification of MDS in 2016, five major subtypes of MDS are renamed (Arber, 2016). More than 40 forms of gene mutations have been identified in MDS; however, 80% of MDS patients have only one or two of these mutations (Harada, 2015). Frequent mutations in MDS include mutations of *SF3B1, EZH2, TET2, SRSF2, ASXL1, RUNX1,* and *TP53.* Notably, most of these mutations can be found also in healthy old people. CCs have low recurrences in MDS (Pellagatti, 2015). Apart from del(5q), frequent forms of CCs in MDS include del(7q), del(20q), (-5), -Y, (+8), and (-7) (Haase, 2008). Some forms of CCs, including (+8), complex karyotype, and abn(7), are associated with poor prognosis of MDS.

### 2.4 Juvenile myelomonocytic leukemia (JMML) is a type of MDS/MPN

JMML is an aggressive form of hematopoietic disorder in infancy characterized by excessive production of monocytes and granulocytes (Toren, 1991; Sethi, 2013). In new WHO classification, JMML is entitled as a type of MDS/MPN (Clara, 2016; Arber, 2016). In JMML, over-production of monocytes and granulocytes leads to extensive infusion of monocytes in organs and in skin. JMML does not have Ph translocation in leukemia cells, and this makes JMML distinguishable from pediatric CML. JMML affects more often boys than girls. JMML is not sensitive to treatment, thus has poor prognosis. Over 20% of JMMLs have abnormality on chromosome 7, such as (-7), (+7), t(7;11), and t(7;20) (de Vries, 2010). The JMML in these cases is also called chromatide-7 syndrome. 30% of JMML children have genetic defects, such as Down syndrome, Shwachman-diamond syndrome, Fanconi anemia, and Kostmann syndrome (Luna-Fineman, 1999). These genetic disorders have all abnormalities on chromosome 7. 90% of JMML patients carry gene mutations of proteins that are related to RAS/MAPK pathway, such as PTPN11, KRAS, NRAS, CBL, and NF1. The frequencies of mutations of *PTPN11, NRAS,* and *NF1* in JMML are respectively 30%, 25%, and 15% (Locatelli, 2015).

### III. Development of myeloid cells (MCs) in bone marrow

AML and CML are both a leukemia resulting from cell transformation of a MC. To understand how a MC is transformed, we firstly make a brief review on the developing process of MCs and the characteristics of MCs. In humans, after birth, all of our blood cells including MCs and lymphoid cells (LCs) are produced by the HSCs in marrow through hematopoiesis. Hematopoiesis has in general four steps: **A.** a HSC produces a progenitor cell; **B.** a progenitor cell then proliferates rapidly to produce a large number of blast cells in different lineages; **C.** a blast cell then proliferates and differentiates into different types of pro-cytes (immature cells); and **D.** a pro-cyte then divides for several times to develop into mature cells (Figure 2). In the present paper, "hematopoietic cell" (HC) is the general name for all the blood forming cells in marrow, including HSCs, MCs, and LCs. Among all the nucleated non-erythroid HCs in marrow, 0.001% are HSCs, 2%-4% are blast cells, 20%-30% are pro-cytes, and more than 60% are mature cells.

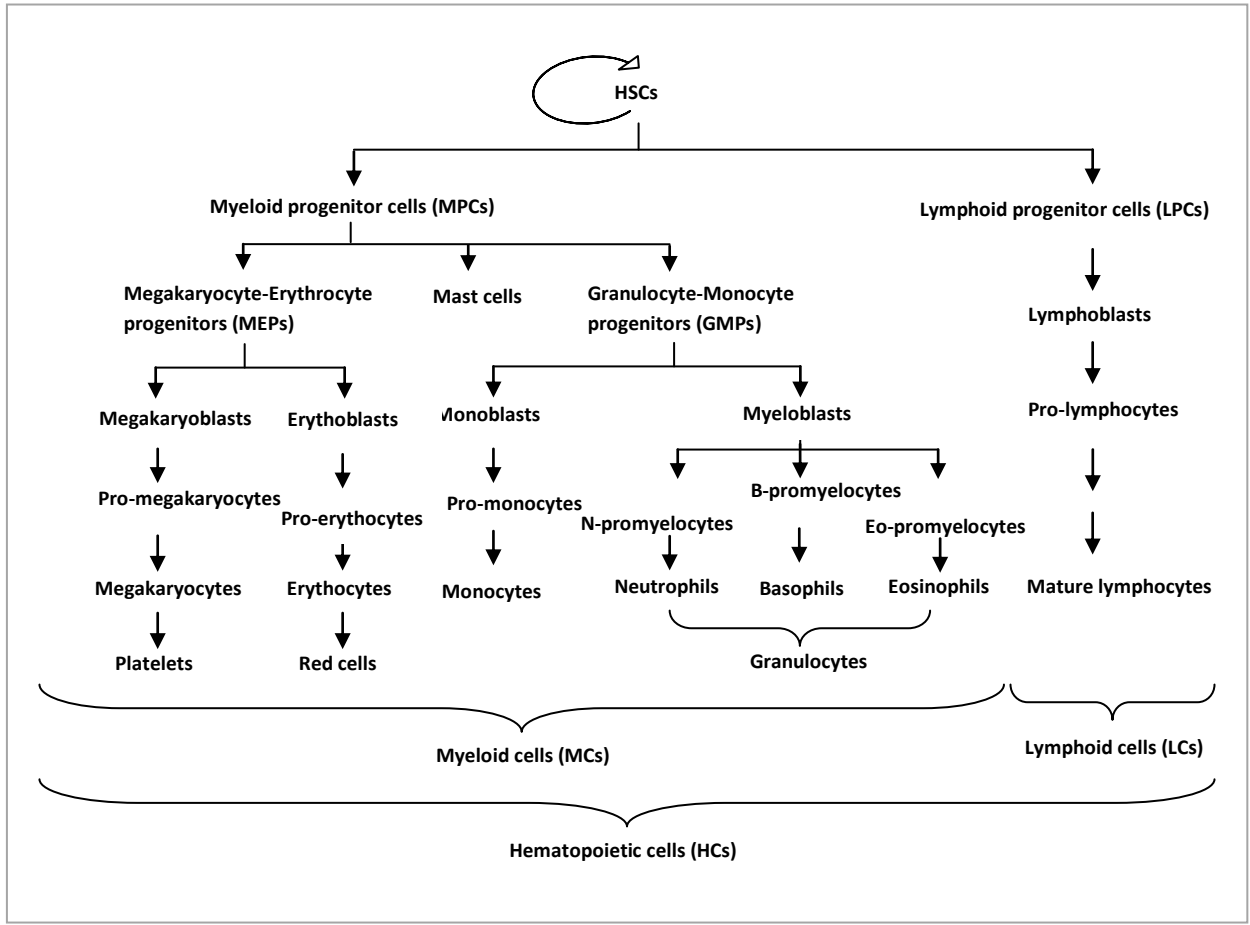

**Figure 2. Development of myeloid cells in marrow**

HSCs are the stem cells for all types of blood cells. For hematopoiesis, a HSC needs to produce a progenitor cell: myeloid progenitor cell (MPC) or lymphoid progenitor cell (LPC). A MPC differentiates firstly into three types of progenitor cells: megakaryocyte-erythrocyte progenitor (MEP), granulocyte-monocyte progenitor (GMP), and mast-cell progenitor. A MEP will differentiate into two types of blast cells: megakaryoblasts and erythroblasts, which will differentiate further respectively into pro-megakaryocytes and pro-erythocytes. A GMP will differentiate into two types of blast cells: monoblasts and myeloblasts. A myeloblast will differentiate further into three types of pro-myelocytes: neutrophilic, basophilic, and eosinophilic, respectively. A monoblast will differentiate uniquely into pro-monocytes. Pro-cytes are also proliferative but uni-potent. Each pro-cyte will differentiate into a number of mature cells of the same type. Myeloid cells (MCs) include all the cells in myeloid lineage, and lymphoid cells (LCs) include all the cells in lymphoid lineage. Hematopoietic cells (HCs) include all the HSCs, MCs and LCs in marrow.

## 3.1 Ancestors of MCs: hematopoietic stem cells (HSCs)

HSCs are the ancestors for all HCs and all blood cells. HSCs are regenerable for our whole lifetime. Through symmetrical cell division, a HSC renews itself by producing two daughter HSCs. By asymmetrical cell division, a HSC produces two different cells: a new HSC and a progenitor cell: myeloid progenitor cell (MPC) or lymphoid progenitor cell (LPC) (Figure 2). MPCs are the ancestors for all the cells of myeloid lineage, LPCs are the ancestors for all the cells of lymphoid lineage. In a normal marrow, 75% of nucleated cells are of myeloid lineage and 20% are of lymphoid lineage. Since HSCs are the unique source of new blood cells, an abnormality on the number or on the functionality of HSCs may lead to a severe disease. For example, aplastic anemia may occur if HSCs have defects on regeneration. If a HSC is transformed by acquired DNA changes, clonal hematopoiesis or failure of hematopoiesis may occur. It is reported that clonal hematopoiesis occurs in 20% of people older than age 90 (Shlush, 2015). Some MDSs may develop from clonal hematopoiesis.

### 3.2 Developing MCs: including progenitor cells, myeloblasts, and pro-cytes

Progenitor cells, myeloblasts, and pro-cytes are precursors of mature MCs. Progenitor cells and blast cells are oligo-potent, whereas pro-cytes are uni-potent (Arai, 2016). A progenitor cell proliferates and differentiates into different types of blast cells, and a blast cell will proliferate and differentiate into one or more types of pro-cytes. For example, a MPC can firstly differentiate into three types of progenitor cells (also blast cells): megakaryocyte-erythrocyte progenitor (MEP), granulocyte-monocyte progenitor (GMP), and mast-cell progenitor. A MEP will differentiate into two types of blast cells: megakaryoblasts and erythroblasts, which will differentiate further respectively into pro-megakaryocytes and pro-erythocytes. A GMP will differentiate into two types of blast cells: monoblasts and myeloblasts. A myeloblast will differentiate further into three types of pro-myelocytes: neutrophilic, basophilic, and eosinophilic, respectively (Figure 2). Differently, a monoblast will differentiate uniquely into pro-monocytes. Pro-cytes are also proliferative but uni-potent. Each pro-cyte will differentiate into a number of mature cells of the same type. In myeloid lineage, a pro-cyte develops into mature cells via three stages of differentiation: pro-cyte, cyte, and meta-cyte. For example, maturation of neutrophilic granulocytes needs to pass three developing stages: pro-myelocyte, myelocyte and mega-myelocyte.

Progenitor cells, blast cells, and pro-cytes are all developing HCs, which are proliferative and immature on cell functions. Blast cells account for 2%-4% of nucleated HCs in marrow. The number of pro-cytes should be much higher than that of blast cells, because a blast cell needs to divide at least one time to develop into pro-cytes. Thus, pro-cytes may represent the largest part of developing HCs in marrow. Although not fully developed, pro-cytes are more mature on cell functions than blast cells. Being large in number and proliferating constantly, developing HCs may have high risk to be injured during DNA synthesis and cell division by damaging factors. In another word, developing HCs have higher risk of DNA injuries and DNA changes than mature cells.

### 3.3 Mature MCs: including granulocytes, monocytes, red cells, and megakaryocytes

A pro-cyte will divide for several times to develop into mature cells. Except T cells, all types of blood cells are fully developed in marrow. Mature MCs have several sub-lineages, including granulocytes (neutrophils, basophils, and eosinophils), monocytes, erythrocytes (red cells), and megakaryocytes (platelets) (Figure 2). Being mature on cell functions and non-proliferative, a mature MC may have lower risk of DNA injuries and DNA changes than a developing MC. Monocytes and granulocytes are the main cells for innate immunity. Monocytes can derive into macrophages, osteoclasts, and dendritic cells. The functions of monocytes are phagocytosis, cytokine production, and antigen-presentation. A monocyte is a large cell with big smooth nucleus and large cytoplasm. This indicates that a monocyte produces a large number of intracellular proteins. A granulocyte is a middle-sized cell, but much bigger than a naïve lymphocyte by its large cytoplasm. A granulocyte produces a great deal of granule proteins in cytoplasm. These proteins are important for the function of a neutrophil: swallowing and digesting bacteria and fungi in cytoplasm.

## IV. A potential source of cell injuries of hematopoietic cells (HCs) in marrow: the repeated bone-remodeling during bone-growth and bone-repair

DNA changes are the driver for cell transformation and for cancer development. However, DNA changes are generated in a cell as consequences of cell injuries and DNA injuries. So far, the main cause for cell injuries of MCs associated with leukemia development is unknown. Some environmental factors such as exposures to radiation or toxic chemicals may be related to leukemia development. However, the high incidence of AML in young children cannot be explained completely by environmental factors. After analyzing the occurring age of ALLs, we have hypothesized that repeated bone-remodeling during bone-growth may be a source of cell injuries of HCs and a trigger for ALL development in children (Wang-Michelitsch, 2018a).

ALL occurs mostly at age 0-20 when the bones are growing. The incidence of ALL after age 25 is very low. This phenomenon suggests that bone-growth may be related to ALL development. In humans, hematopoiesis takes place mainly in marrow cavities and the spongy part of bones. Marrow cavity and spongy bone are developed and enlarged with the growth of a bone. Bone-growth is a result of repeated modeling-remodeling of bone tissues (Rockville, 2004). Bone-remodeling is a process of digestion of the bone tissues exposed to marrow cavity by osteoclasts. Osteoclasts digest bone-matrix by secreting acid substances and proteinases. It is quite possible that these digestive substances and the bone debris produced during bone-remodeling injure occasionally the HCs in marrow cavity. Naturally, the risk of a HC to be injured by bone-remodeling is quite low. However, long-term (25 years) repetition of bone modeling-remodeling for bone-growth can highly increase this risk.

In addition, young children have high incidence of bone fractures because of their fragilities on muscles and bones. Bone injuries may disturb bone-growth and increase the frequency of bone-remodeling. Thus, in a child, bone injuries (external) increase the risk of cell injuries of HCs by bone-remodeling during bone-growth (internal). In fact, bone-remodeling occurs also during bone-repair and bone adaptation. Thus HCs can be affected by bone-remodeling at all ages. Bone-remodeling may be a causing factor not only for pediatric leukemia but also for adult leukemia.

Cell injuries by internal damage may occur also to other cells including tissues during body development and during inflammations. DNA changes can be also generated in some injured tissue cells. However, for a tissue cell, cell transformation needs a long time as a co-effect of external and internal factors. Differently, a MC can be transformed more rapidly than a tissue cell as that seen in AML and CML. Thus, the effect of internal damage on a MC can be recognized. We will discuss in next parts why a MC can be transformed more rapidly than a tissue cell.

## V. Generation and accumulation of DNA changes in HCs is a result of repeated cell injuries and repeated cell proliferation

There are two major types of DNA changes: point DNA mutation (PDM) and chromosome change (CC). PDM appears as alteration, deletion, or insertion of one or two bases in a DNA. PDM is called also gene mutation and molecular mutation. CC is called also cytogenetic change and abnormal karyotype. A CC can be a numerical CC (NCC) or a structural CC (SCC). A NCC exhibits as loss or gain of one or more chromosomes. For example, (-7), (+7), hyperdiploid, and aneuploid are all forms of NCCs. A SCC exhibits as rearrangement of part of a chromosome, such as translocation (t), deletion (del), gain (+), inversion (inv), and amplification (amp) of part of a chromosome.

Different types of DNA changes are generated in a cell by different mechanisms. Studies show that a PDM is generated as a result of Misrepair of DNA on a double-strand DNA break (DSB) (Rothkamm, 2002; Natarajan, 1993; Bishay, 2001). Similarly, generation of SCC is also a result of Misrepair of DNA. A SCC can be generated when a cell has multiple DSBs (Wang-Michelitsch, 2018a). Differently, generation of NCC is rather a consequence of dysfunction of cell division promoted by damage. In our view, a PDM/SCC is made for repair of DNA for preventing cell death, thus it is not really a mistake. Differently, a NCC is not generated for "repair"; thus survival of a cell with a NCC is a real mistake. Unfortunately, a NCC can affect multiple genes and cause cell transformation rapidly.

Normally, a somatic cell has quite low opportunity to survive from a DSB by Misrepair of DNA. Thus, DNA mutations cannot accumulate in "the same" cell, but can only accumulate in its offspring cells (Wang-Michelitsch, 2015). Cell injuries are the triggers for generation of DNA changes. Cell proliferation enables the accumulation of DNA changes generated in different generations of offspring cells. Thus, accumulation of DNA changes in cells is a result of repeated cell injuries and repeated cell proliferation. For example, in bone marrow, repeated cell injuries of HCs may trigger generation of DNA changes in some HCs. Regeneration of HSCs and proliferation of HCs enable the DNA changes to accumulate in these cells. However, long-term accumulation of DNA changes (longer than months) can occur only in HSCs. Namely, only the DNA changes that are generated or inherited in HSCs can accumulate for a long time. But all the offspring cells of a HSC, including MCs, LCs, and mature blood cells, can inherit the DNA changes accumulated in the HSC. For example, a MC can have all the DNA changes of its precursor HSCs and other precursor cells (Wang-Michelitsch, 2018a).

## VI. Different types of DNA changes have different effects on a cell

DNA changes are the driver for cell transformation; however, not all the DNA changes contribute to cell transformation. Big CCs may cause cell death, but silent PDMs have no effect on a cell. CCs are rare in cancers developed from tissue cells, but not rare in leukemia and lymphoma. This phenomenon indicates that a tissue cell can be transformed mainly by PDMs (gene mutations), but a blood cell and a HC can be transformed not only by PDMs but also possibly by CCs.

### 6.1 Effect of a point DNA mutation (PDM) on a cell

A PDM can affect at most "one" gene. Since most regions of a DNA (in one chromosome) are non-coding for proteins, most PDMs are silent in a cell. However, if a PDM occurs in the coding region of a gene and affects expression of the gene, the PDM is not silent. There may be two situations: **A.** if the gene is essential for cell survival and it is expressed constitutively, production of non-functional proteins or failure/reduction of gene expression can make the cell die; and **B.** if the gene is expressed only when there is an inducer, failure of gene expression does not necessarily cause cell death. In situation **A**, the PDM is fatal, but in situation **B**, the PDM is mild. Hence, most PDMs are silent or mild, by which they can "survive" and accumulate in cells. By accumulation, some mild PDMs may exhibit their "emergent effects" on a cell. If some of the PDMs make a cell able to proliferate independently, the cell is transformed. It is possible that some forms of PDMs do not affect directly cell functions but they may alter genome stability and accelerate generation of other DNA changes. These PDMs also contribute to cell transformation.

### 6.2 Three types of chromosome changes (CCs) by their effects on a cell

In AML, CCs can be found in 80% of pediatric cases but only 45% of adult cases. This phenomenon suggests that the rapider cell transformation of a MC in pediatric AML is related to CCs. In general, a CC has greater effect than a PDM on a cell, but different forms of CCs may have different effects. Thus it is essential to regroup CCs by their impacts on a cell. We classify CCs into three groups by their effects on a MC: great-effect CCs (GECCs), mild-effect CCs (MECCs), and intermediate-effect CCs (IECCs) (Box 1).

#### 6.2.1 *Great-effect CCs (GECCs)*

GECC is a type of CC that affects one or more genes and can alone drive cell transformation in "one step" (Box 1). For most types of cells, a GECC will cause cell death. However, for certain types of cells, a GECC may be not fatal, but rather be sufficient for driving cell transformation directly (Wang-Michelitsch, 2018b). The aneuploid in pediatric ALL and the t(8;14) in Burkitt lymphoma (BL) are two examples of GECCs. ALL and BL have both a peak incidence in children, indicating that a lymphoblast and a centroblast have a risk to be transformed by a GECC. Differently, AML and CML do not have a peak incidence in children and other ages. This suggests that a MC cannot be transformed in "one-step" by a GECC. In another word, a MC cannot survive from a GECC.

**Box 1. Different types of DNA changes have different effects on a cell**

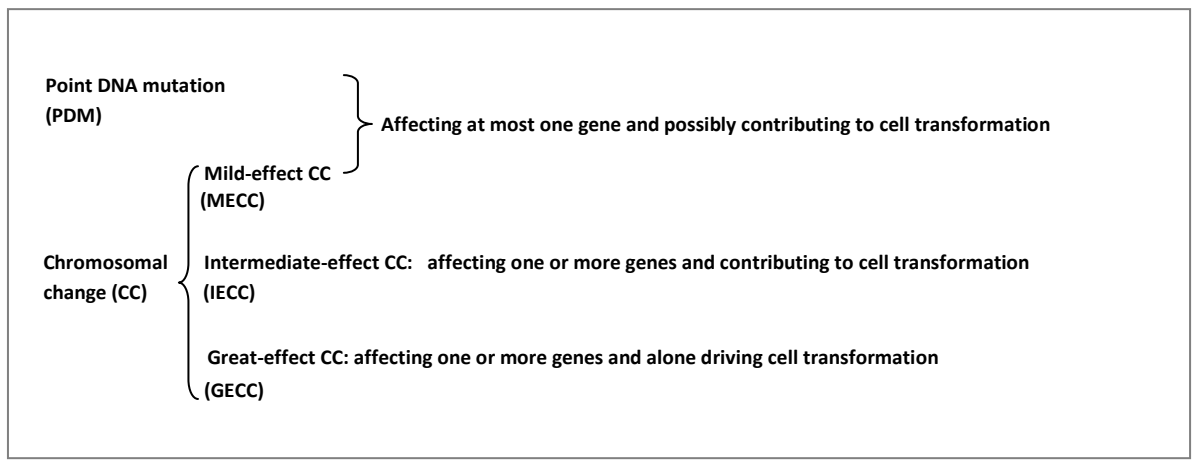

### 6.2.2 Mild-effect CCs (MECCs)

MECC is a type of CC that has silent or mild effect on a cell (Box 1). A MECC can be a CC occurred to a non-coding region of a DNA or to a coding part of a DNA but not affecting gene expression. Thus, a MECC has similar effect to a PDM on a cell. Among all the DNA changes in AML, those that are associated with low-risk prognosis, including t(15; 17), t(8; 21), and inv(16), may be MECCs. Most of MECCs can accumulate in cells, and some of them may contribute to cell transformation. Since a NCC (numerical CC) can affect multiple genes, MECCs are often tiny SCCs (structural CCs).

### 6.2.3 Intermediate-effect CCs (IECCs)

IECC is a type of CC that affects one or more genes and contributes to cell transformation (Box 1). An IECC cannot alone cause cell death or cell transformation. A mature MC and a HSC may have low tolerance to an IECC. Thus, an IECC is generated more often in a developing MC, such as progenitor cell, blast cell, and pro-cyte. A cell may have increased risk of generation of IECC, when it has already other DNA changes. Some PDMs/MECCs may affect the chromosome stability in a cell and increase the sensitivity of a DNA to damage. Namely, the more PDMs/MECCs a cell has, the more risk has the cell on generation of IECC. Thus, IECCs have increasing occurrence in MCs with age of our body. IECCs may also accumulate in MCs, but this opportunity is low.

The cell transformation driven by an IECC is a co-effect of the IECC and other PDMs/MECCs. One reason is that: generation of IECC is often related to early PDMs/MECCs in the cell. With stronger effect than a PDM/MECC, an IECC can accelerate cell transformation. An IECC drives cell transformation possibly by affecting an oncogene or a tumor suppressing gene. A good example of IECC is the Ph translocation (t(9;22)) in CML. Ph translocation contributes to the cell transformation of MC in CML by generating fusion gene of *BCR-ABL1* and activating permanently ABL1 tyrosine kinase. CML occurs rarely in young children, but it has increasing incidence with age (Kang, 2016). This phenomenon suggests that Ph translocation has increasing occurrence in MCs with age of an individual. IECCs might represent the cytogenetic abnormalities that are associated with intermediate- and high-risk prognosis of AML. These CCs include t(9; 11), (+8), (-y), t(6;9), t(9;22), (-5), (-7), inv (3), and abn(3q). An IECC is a driver DNA change, thus it can be chosen to be a blocking target in targeted cancer treatment.

### VII.  A MC may have higher survivability from DNA changes than a tissue cell

Different forms of DNA changes have different effects on a cell; however, the same form of DNA change may have different effects on different types of cells. MCs at different developing stages may have also different tolerances to a DNA change. In this and next parts, we discuss why a MC can be transformed more rapidly than a tissue cell. In our view, two factors make it possible that MC is transformed more rapidly. These two factors are: **A.** a MC may have higher tolerance to DNA changes than a tissue cell; and **B.** A MC may require obtaining fewer cancerous properties for transformation than a tissue cell.

### 7.1  Higher tolerance of a MC to DNA changes than a tissue cell

A blood cell is different from a tissue cell by surviving environment. Here, "blood cells" include all the mature blood cells and all the hematopoietic cells in bloodstream and in marrow. A tissue cell is anchor-dependent for survival, but a blood cell is anchor-independent. If a tissue cell fails to anchor to its neighbor cells or extracellular matrixes (ECMs), the cell will die through apoptosis (cell suicide). For a tissue cell, cell adhesion molecules are essential for cell survival. Major types of adhesion molecules include cadherins, selectins, and integrins (Li, 2012). If a DNA change affects production of one of these molecules in a tissue cell, the cell may die. A blood cell expresses also adhesion molecules. However, for a blood cell, the adhesion molecules are used for cell development and cell functions, but not used for anchoring to other cells/ECMs for survival. For example, a mature granulocyte can survive no matter whether it is in bone marrow, bloodstream, or tissue. Thus, cell adhesion molecules of a granulocyte are not essential for cell survival. Failure of expression of these molecules by DNA changes does not necessarily cause death of a granulocyte.

In addition, at different developing stages, MCs express different CD molecules. For example, CD33 and CD34 are only expressed on developing MCs but not on mature MCs (Dutta, 2014). In contrast, molecules including CD17 and CD93 are only expressed on mature MCs. MPCs (myeloid progenitor cells) express CD34, CD33, and CD38, but GMPs express CD33, CD34, and CD31. Myeloblasts produce CD13, CD15, and CD33, whereas mature neutrophils produce CD17, CD32, CD35, CD43, and CD93. These molecules are not expressed permanently but rather by inducers in environment. Inducible expression of CD molecules implies that a MC can survive no matter these molecules are expressed or not. If a mutation occurs to the gene of one of these molecules, the mutation may not affect the cell. Taken together, due to the anchor-independence for survival and the inducible expression of cell surface molecules, a MC and other blood cells may have higher survivability from DNA changes than a tissue cell. Namely, a DNA change that is fatal for a tissue cell may not be necessarily fatal for a MC.

## 7.2 A developing MC may have higher survivability from DNA changes than a mature MC

A developing MC such as progenitor cell, myeloblast, and pro-cyte is immature on cell functions, thus it may be more tolerant to DNA changes than a mature MC. Having high integrity on cell functions, a mature cell is alert to any tiny change occurred to the cell. However, because of immaturity on some cell functions, an immature MC may be unable to "sense" a change in the cell. In addition, to have full cellular functionality, a mature cell needs to express more types of molecules than an immature one. Thus, more genes should be activated in a mature cell. A DNA change that occurs to a gene that is activated only in mature cells can only affect a mature cell but not a developing cell. Namely, a DNA change that is fatal for a mature MC may be not fatal for an immature MC. Therefore, a developing MC may have higher tolerance to DNA changes than a mature cell. The high survivability from DNA changes makes a developing MC have a risk to be transformed by a CC (chromosome change).

However, a developing MC may have lower tolerance to DNA changes than a developing lymphoid cell (LC), such as lymphoblast and pro-lymphocyte. It is known that naïve lymphocyte is the smallest cell in our body. A naïve lymphocyte has little cytoplasm, indicating that it does not produce many intracellular proteins. This means that most genes may be switched-off in a naïve lymphocyte and its precursor LCs. Differently, a mature MC such as monocyte and granulocyte has a large cytoplasm. This indicates that a monocyte/granulocyte produces a great deal of intracellular proteins. It is probable that many genes that are not activated in a LC are activated in a MC. Thus, a DNA change that has an effect on a MC may have no effect on a LC. Namely, a LC may be tolerant to more types and more number of DNA changes than a MC. Taken together, with respect to the tolerance to DNA changes, developing LC > developing MC > mature MC > tissue cell.

## VIII. A MC may require obtaining fewer cancerous properties for transformation than a tissue cell

A tumor can grow rapidly and even invade into other organs due to some cancerous properties of the tumor cells (Cooper, 2000). Among all cancerous properties, the most important five are: stimulator-independent mitosis (SIM), loss of cell-contact inhibition (LCCI), production of matrix metalloproteinase (PMM), anoikis-resistance (AR), and acquired mobility (AM) (Table 2). SIM and LCCI are two essential properties for a non-malignant tumor cell. SIM gives a cell the ability of stimulator-independent proliferation. LCCI enables an unlimited cell proliferation by overcoming the cell-contact inhibition in a tissue. PMM, AR, and AM are essential properties for a malignant tumor cell. These three properties make a tumor cell invasive and metastatic. For a tissue cell, SIM and LCCI are the first two properties to acquire for cell transformation. Thus, transformation of a tissue cell often begins by non-malignant transformation. When a tumor cell acquires additionally one of PMM, AR, and AM, it undergoes malignant transformation (Wang-Michelitsch, 2018b).

Interestingly, a blood cell and a HC are born with some properties similar to that of a cancer cell. Firstly, a blood cell is anoikis-resistant, thus it is AR-positive. Secondly, cell proliferation of HCs in marrow is not restricted by cell-contact inhibition; hence HCs are LCCI-positive. Thirdly, a blood cell has auto-mobility for migration in bloodstream, lymph, and tissues, thus it is AM-positive. Finally, a blood cell can produce matrix metalloproteinases during migration, thus it is PMM-positive. Taken together, a blood cell/HC is born with four properties that are needed for cell transformation, and these properties are LCCL, PMM, AR, and AM (Table 2). Hence, a blood cell/HC needs to obtain only "one" property: namely the SIM, for transformation. This means that a blood cell/HC can be transformed by fewer number of DNA changes than a tissue cell, thus it can be transformed more rapidly.

Table 2. A blood cell may require obtaining fewer cancerous properties for transformation than a tissue cell

| Cancerous properties | Properties required to obtain for cell transformation | | |
|---|---|---|---|
| | | Tissue cell | Blood cell and hematopoietic cell |
| Stimulator-independent mitosis (SIM) | **For non-malignant transformation** | 1 | 1 |
| Loss of cell-contact inhibition(LCCI) | | 1 | 0 |
| Production of matrix metalloproteinase (PMM) | **For further malignant transformation** | 1 | 0 |
| Anoikis-resistance (AR) | | 1 | 0 |
| Acquired Mobility (AM) | | 1 | 0 |

## IX. Our hypothesis: a MC may have two pathways on transformation: a slow and an accelerated

Blood cancers including leukemia and lymphoma have high frequency of CCs in cancer cells. This indicates that CCs contribute to the cell transformations of MCs and LCs in blood cancers. As a type of blood cell, a MC may have higher survivability from DNA changes and require obtaining fewer cancerous properties for transformation than a tissue cell. On this basis, we hypothesize that *a MC may have two pathways on transformation: a slow and an accelerated* (Box 2). Slow pathway is driven by accumulation of PDMs and MECCs through many generations of cells. Accelerated pathway is driven by accumulation of PDMs, MECCs, and IECC(s) through a few generations of cells. Transformations via different pathways occur at different ages. Thus, distinguishing between these two pathways of transformation of a MC can help us understand the age-specific incidences in AML, CML, and MDS.

**Box 2. Hypothesis: a myeloid cell may have two pathways on cell transformation**

- **Slow pathway**: by accumulation of PDMs and MECCs through many generations of cells

- **Accelerated pathway**: by accumulation of PDMs, MECCs, and IECC(s) through a few generations of cells

PDM: point DNA mutation
MECC: mild-effect chromosome change
IECC: intermediate-effect chromosome change

## 9.1 Slow pathway: by accumulation of PDMs and MECCs through many generations of cells

A MC can be transformed when it acquires the property of stimulator-independent mitosis (SIM) by DNA changes. As discussed earlier, a normal MC including developing MC cannot survive from a GECC. Thus, a MC can acquire SIM only in two ways: by accumulation of PDMs/MECCs or by accumulation of PDMs/MECCs and IECC(s). A cell transformation driven by accumulation of PDMs/MECCs can occur to any one of MCs. In such a transformation, the last driver PDM/MECC is generated in the first transformed cell, but other driver PDMs/MECCs are generated in precursor HSCs/MCs but inherited by the transformed cell. Since HSCs are the unique long-living stem cells in marrow, long-term accumulation of PDMs/MECCs takes place mainly in HSCs. The accumulation of PDMs/MECCs in HSCs is a result of repeated cell injuries and repeated regenerations of HSCs. Accumulation of PDMs and MECCs is a slow process, because it needs to proceed through many generations of HSCs. Hence, the pathway of transformation by accumulation of PDMs/MECCs is a "slow pathway" (Figure 3).

As a result of repetition of cell injuries, the incidence of cell transformation via this pathway increases with age. Thus, cell transformation via slow pathway takes place mainly at old age. A leukemia that results from cell transformation via slow pathway occurs mainly in adults and rarely in children. Thus, MDS may develop mainly via this pathway (Figure 4). Some adult subtypes of AML, such as **M6**, may develop via this pathway. CML occurs at any age; hence it may not develop via this pathway.

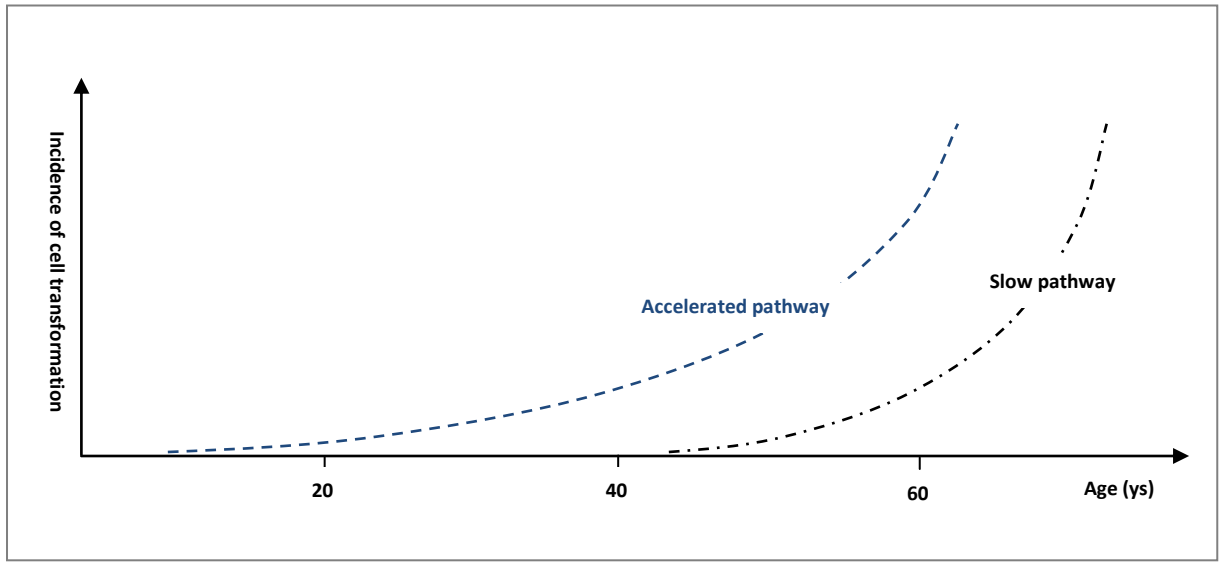

**Figure 3. Transformations of a MC via different pathways occur at different ages (a schematic graph)**

A MC may have two pathways on transformation: a slow (**Black line**) and an accelerated (**Blue line**). Slow pathway is driven by accumulation of PDMs and MECCs through many generations of cells. Accelerated pathway is driven by accumulation of PDMs, MECCs, and IECC(s) through a few generations of cells. Transformations via different pathways occur at different ages. A cell transformation via slow pathway occurs mainly in adults and has increasing incidence with age. A cell transformation via accelerated pathway can occur at any age and has also increasing incidence with age.

## 9.2 Accelerated pathway: by accumulation of PDMs, MECCs, and IECC(s) through a few generations of cells

A MC can obtain SIM also by accumulation of PDMs, MECCs, and IECC(s). Some forms of CCs in AML, such as t(9; 11), (+8), (-y), t(6;9), t(9;22), (-5), (-7), inv (3), and abn(3q), may be IECCs. With stronger effect than a PDM/MECC, an IECC can accelerate cell transformation driven by accumulation of PDMs/MECCs. Thus, a cell transformation driven by IECC(s) and PDMs/MECCs is rapider than that driven only by PDMs/MECCs in slow pathway. We refer the IECC-driven pathway as to an "accelerated pathway" (Figure 3). A transformation via accelerated pathway needs to be achieved also through some generations of cells, but fewer than that in slow pathway. An IECC can be generated in a cell at any time; however it may have increasing occurrence in MCs with age of our body. Therefore, a cell transformation via accelerated pathway can occur at any age, but it has increasing incidence with age. A leukemia that develops via this pathway occurs in both adults and children, but more in adults. Thus, CML develops more likely via this pathway, and Ph translocation is the IECC in CML (Figure 4). Pediatric AMLs and some adult AMLs, such as **M1, M2, M3, M4,** and **M5**, may develop via this pathway.

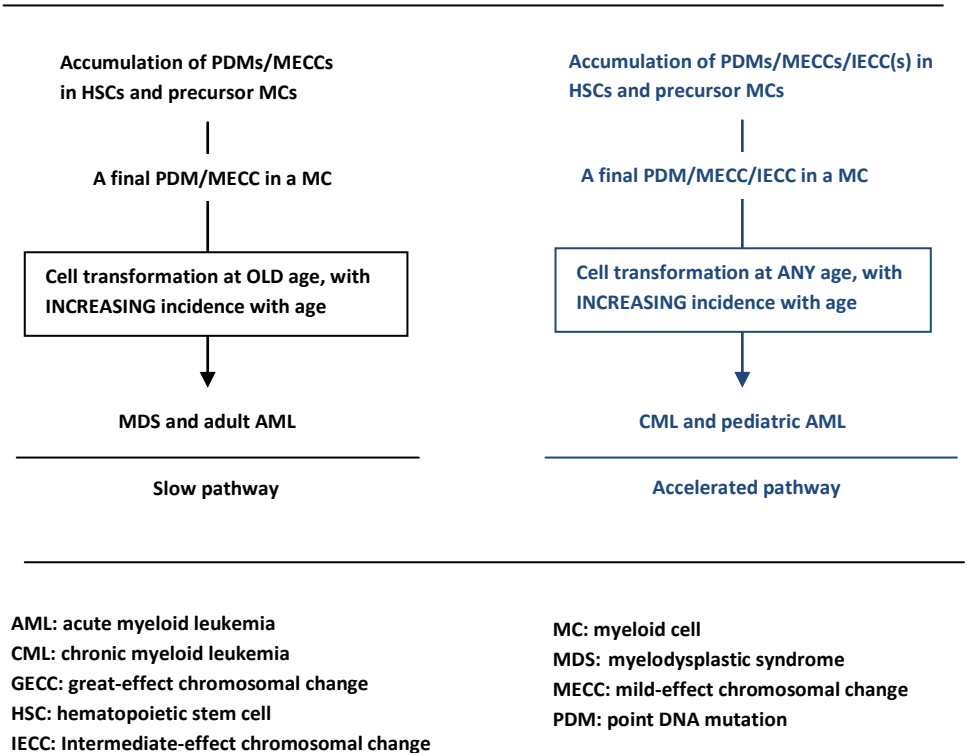

**Figure 4. The age of occurrence of leukemia is determined by the transforming pathway of a MC**

A MC may have two pathways on cell transformation: a **slow pathway** by accumulation of PDMs and MECCs through many generations of cells, and a**n accelerated pathway** by accumulation of PDMs, MECCs, and IECC(s) through a few generations of cells. In these two pathways, the final PDM/MECC/IECC is generated in the first transformed cell, but other PDMs/MECCs/IECC(s) are generated in precursor HSCs/MCs. Transformations via different pathways occur at different ages. Thus, the age of occurrence of leukemia is determined by the transforming pathway of a MC. A leukemia that occurs mainly in adults, such as MDS and adult AML, is often a result of cell transformation via slow pathway. A leukemia that occurs in both children and adults, such as CML and pediatric AML, is more likely a result of cell transformation via accelerated pathway.

Similar to that in slow pathway, most of the driver PDMs/MECCs in a cell transformation via accelerated pathway are generated in the precursor HSCs/MCs of the first transformed cell. Thus, cell injuries of HSCs in marrow may contribute to developments of all forms of myeloid leukemia, including AML, CML, and MDS. Since HSCs have low tolerance to an IECC, the IECC associated with development of AML/CML may be more often generated in the first transformed MC or one of its precursor MCs.

## X.   Three grades of transformation of a MC: low, high, and intermediate

Leukemia is a fatal disease, because accumulation of leukemia cells in marrow cavity can cause failure of hematopoiesis. Failure of cell differentiation of leukemia cells is the main reason for the accumulation of leukemia cells in marrow cavity. Thus, the severity of leukemia is determined mainly by the degree of cell differentiation of leukemia cells. If the cell differentiation is normal, leukemia cells can develop into mature cells, which can enter bloodstream and do not accumulate in marrow. However, if the cell differentiation is abnormal, leukemia cells cannot develop into mature cells and have to accumulate in marrow.

Thus, the cell transformations of a MC in different forms of leukemia may be different by "grades". Transformation of a cell can be differentiation-affected or differentiation-non-affected. If cell differentiation is not at all affected, the transformation is at **low-grade**. If cell differentiation is severely affected, the transformation is at **high-grade.** There is still a third situation: if cell differentiation is partially affected, the transformation is at **intermediate-grade.** Thus, a cell may have three grades of transformation: low-grade, high-grade, and intermediate-grade. A tissue cell is often transformed firstly at low-grade then high-grade. However, for a MC, the situation is different. A developing MC can be transformed directly at all three grades, but a mature MC needs to be transformed firstly at low-grade.

In a cancer developed from a tissue cell, the grade of transformation determines whether the cancer is non-malignant (low-grade) or malignant (intermediate-grade or high-grade). However, for a MC, transformations at different grades will lead to different forms of leukemia. For example, low-grade transformation of a MPC may result in CML development; however transformation of a MPC at intermediate-grade or high-grade may result in MDS development (Table 3). Low-grade transformation of a developing MC may result in MPN development; however transformation of a developing MC at intermediate-grade or high-grade may result in AML development. In both of slow and accelerated pathways, the cell transformation of a MC can be firstly at low grade, leading to occurrence of a proliferative disease such as CML and MPN. However, when one of the leukemia cells in CML/MPN is transformed further at intermediate-grade or high-grade, MDS/AML may occur. Low-grade transformation of a HSC may result in occurrence of clonal hematopoiesis.

Table 3. The grades of cell transformation of a MC in AML, CML, MDS, and MPN

| Cell of origin | Low-grade cell transformation | Intermediate-grade or high-grade cell transformation |
| --- | --- | --- |
| HSC | Clonal hematopoiesis | MDS? |
| MPC | CML | MDS |
| Developing MC | MPN | AML |

## XI. Developments of pediatric AML and CML may be associated with the repeated bone-remodeling during bone-growth

Our analysis shows that pediatric AML/CML may be a result of transformation of a MC via accelerated pathway. In this part, we will interpret the developing characters of AML, CML, and MDS by our hypothesis. Firstly, although these three forms of leukemia are different by occurring age, pathology, and prognosis, they may have common causes. They may be all related to the cell injuries of HSCs and MCs caused by external and/or internal damaging factors. For adult cases, repeated exposures to radiation/chemicals and repeated bone-remodeling during bone-growth and bone-repair may be all related to leukemia development. However, for pediatric cases, repeated bone-remodeling during bone-growth may be a more important causing factor. On all three forms of myeloid leukemia, men have higher incidences than women. One reason may be that: men have higher risk of bone injuries by their heavier body weight and heavier physical work.

**11.1 AML: as a result of high-grade transformation of a developing MC via slow or accelerated pathway**

AML can occur at any age and has increasing incidence with age. Thus, AML may develop as a result of cell transformation of a MC via slow or accelerated pathway. AML has two times of incidence as CML. The higher incidence of AML than CML may be due to two factors: **A.** AML can develop from all types of developing MCs, but CML arises only from a MPC; and **B.** AML can develop via both slow and accelerated pathways, but CML develops mainly via accelerated pathway. Pediatric AML develops more likely via accelerated pathway, whereas adult AML may develop via slow or accelerated pathway (Table 4). Among all the CCs in AMLs, some forms including t(6; 9), (–5), inv (3), t(9;22), and abn(3q) may be IECCs, because they are associated with a poor prognosis of AML. The AMLs that have these IECCs may develop via accelerated pathway. Other forms including t(15; 17), t(8; 21), inv(16), t(11;17), and t(9;11) may be MECCs, because they are related to a good prognosis of AML.

The severity of an AML is determined by two factors: the maturity of the cell-of-origin and the grade of cell transformation. For a developing MC, high-grade transformation will lead to failure of cell differentiation and production of immature leukemia cells; low-grade transformation will lead to over-production of mature cells; and intermediate-grade transformation may result in insufficient production of mature cells. The hematopoiesis is affected in all subtypes of AML, suggesting that the cell transformations of MCs in AMLs are all at high-grade or intermediate-grade. Among all subtypes, **M3**, **M4**, and **M5** have better prognoses than others, suggesting that the leukemia cells in these subtypes can produce mature cells. Thus, the cell transformations in **M3, M4,** and **M5** may be all at intermediate-grade. Similarly, in **M1** and **M2,** the leukemia cells have minimal or partial maturation, suggesting that **M1** and **M2** may be results of intermediate-grade cell transformations of MCs.

On this basis, we make a prediction for the cell of origin and the grade of cell transformation in each subtype of AML: **M0**: by high-grade transformation of a myeloblast; **M1** and **M2**: by intermediate-grade transformation of a myeloblast; **M3**: by intermediate-grade transformation of a pro-myelocyte; **M4** and **M4eos**: by intermediate-grade transformation of a GMP (Granulocyte-Monocyte Progenitor); **M5**: by intermediate-grade transformation of a pro-monocyte; **M6**: by high-grade or intermediate-grade transformation of an erythroblast or a pro-erythrocyte; and **M7**: by high-grade or intermediate-grade transformation of a megakaryoblast or a pro-megakaryocyte. Although AML develops by high-grade or intermediate-grade transformation of a developing MC, it may begin sometimes by low-grade transformation, which may appear as an asymptomatic MPN.

Table 4. Pathways of cell transformation of a MC in AML, CML, and MDS

|  | Driver DNA changes | Cell of origin | Pathway of transformation | Grade of transformation |
|---|---|---|---|---|
| **Pediatric AMLs and some adult AMLs** | Accumulation of PDMs/MECCs/IECC(s) | Developing MC | Accelerated | High or intermediate |
| **Adult subtype of AML** | Accumulation of PDMs/MECCs | Developing MC | Slow | High or intermediate |
| **CML** | Accumulation of PDMs/MECCs + Ph translocation | MPC | Accelerated | Low |
| **MDS** | Accumulation of PDMs/MECCs | MPC/HSC? | Slow | High or intermediate |

## 11.2 CML: as a result of low-grade transformation of a myeloid progenitor cell (MPC) via accelerated pathway

In CML, all the lineages of MCs are affected, thus CML originates more likely from a MPC (Table 4). Since HSCs have low survivability from an IECC, the Ph translocation in CML may be generated in a MPC. MPCs can regenerate for some generations. Ph translocation is found in 95% of CMLs but only in 5% of AMLs. This suggests that a MPC may have higher tolerance to Ph translocation than other developing MCs. However, Ph translocation is the main form of CC in CML-CP, indicating that a MPC may have low survivability from a CC except Ph translocation. In another word, apart from Ph translocation, a MPC can be transformed mainly by accumulation of PDMs/MECCs.

CML is a proliferative disease, suggesting that it is a result of low-grade transformation of a MPC (Table 3). The progression of CML from chronic phase to blast crisis may be a result of cell transformation of a MPC from low-grade to high-grade. The CCs observed in CML, including t(8;21), second Ph translocation, iso(17q), and (+19), may be secondary DNA changes. However, some of them may drive further transformation of CML cells. Ph translocation may be related to the generation of some of the secondary DNA changes. As a form of CC, Ph translocation may affect the genomic stability in a cell and make a CML cell more fragile to DNA damage (Neviani, 2014).

CML is a form of MPN. However, CML can occur at all ages but other types of MPNs such as ET and PV occur only in adults. Other types of MPNs do not have Ph translocation in tumor cells. This suggests that Ph translocation is an accelerating factor in CML development. Namely, it is the Ph translocation that is responsible for the occurrence of CML in a child. TKIs are effective drugs in controlling CML. However, TKIs can only inhibit cell proliferation but cannot kill leukemia cells. Thus, for a CML patient, TKIs need to be given for a long period of time, and this is very expensive.

## 11.3  MDS: probably as a result of high-grade transformation of a MPC via slow pathway

Occurring mainly in old people, MDS develops more likely via slow-pathway of cell transformation (Table 4) (Harada, 2015). MDS originates more likely from a MPC. The reason is that: in MDS, the hematopoiesis of all the myeloid lineages is affected, but that of lymphoid lineage is not affected. The defective hematopoiesis in MDS should be a consequence of failure of cell differentiation of the leukemia cells. This indicates that MDS develops by high-grade or intermediate-grade transformation of a MPC. Ph translocation is the main form of CC in CML-CP. This indicates that a MPC may have low tolerance to a CC except Ph translocation. Hence, the cell transformation of a MPC in MDS development is driven mainly by accumulation of PDMs/MECCs. MDS may begin by low-grade cell transformation of a HSC/MPC, which appears as clonal hematopoiesis. Studies show that some MDS-associated gene mutations can be found in blood cells in 5% of healthy old people (Harada, 2015). These people may have clonal hematopoiesis indeterminate potential (CHIP). CHIP may proceed into MDS when an offspring MPC of the clonal HSCs undergoes intermediate-grade or high-grade transformation. Although forms of CCs are observed in MDS cells, including del(5q), del(7q), del(20q), (-5), and (-7), they may be mostly secondary DNA changes.

MDS and secondary AML are often treatment-resistant. Two factors may be associated with the treatment-resistance of an adult cancer (leukemia): the heterogeneity of cancer cells and the existence of pre-cancer cells. Firstly, cancer heterogeneity is a result of generations of secondary DNA changes in cancer cells. During the slow progression of an adult cancer, different secondary DNA changes may be generated in different cells. Thus, different cancer cells will have different sensitivities to a treatment. Secondly, in an adult cancer, the "sister cell" and some of "cousin cells" of the first transformed cell are s kind of "pre-cancer cells". These cells have most of the DNA changes that the transformed cell has, because they all have the same precursor HSCs and MCs. A pre-cancer cell can be promoted to transform by chemotherapy or radiotherapy (Rothkamm 2002). Thus, existence of pre-cancer cells may be a reason for the relapse of an adult cancer (leukemia). In targeted treatment, the selected target DNA changes should be those that are generated earlier, because these DNA changes may exist in all tumor cells. However, further progression of a cancer may anyway alter the sensitivities of cancer cells to a treatment. Targeted treatment may be beneficial for some patients, but its effect is limited.

### 11.4 Infant AML and JMML: are they consequences of defective development of MCs?

AML has a small peak incidence before age 2. In our view, only a cell transformation via rapid pathway can have a peak incidence at certain age (Wang-Michelitsch, 2018b). Thus, we hypothesize that some Infant AMLs (IAMLs) may develop by rapid cell transformation of an abnormal-developed MC by a great-effect CC (GECC). In our recent discussions on lymphoid leukemia, we have proposed that a developing lymphoid cell such as lymphoblast may have a risk to be transformed by a GECC. We found out that, by the hypothesis of a rapid pathway of transformation of a lymphoblast, we can understand the big peak incidence of ALL in children (Wang-Michelitsch, 2018b). However, a normal developing MC may be not tolerant to a GECC and unable to be transformed via rapid pathway. In addition, ALL peaks at age 2-5, but IAML peaks at age 0-2. The peak of IAML is much smaller than that of ALL. These differences between ALL and IAML indicate that IAML may have a distinct causing factor from pediatric ALL.

It is known that IAML children have often hereditary conditions, including Down syndrome, Fanconi anemia, Diamond-Blackfan syndrome, and Li-Fraumeni syndrome (Saida, 2017). A child with Down syndrome has 16-fold risk higher than other children on IAML development. It is possible that infants of these syndromes have deficiencies on functions of HSCs/MCs due to congenital genetic abnormalities. A genetic abnormality may have two effects on a MC: defective cell differentiation and reduced genome stability. On one hand, the defective cell differentiation can make a MC have increased tolerance to DNA changes. On the other hand, forms of CCs such as t(1;22) and *MLL* rearrangements can be generated in a MC as a consequence of genome instability. As a co-effect, a poor-developed MC may be able to survive from a GECC and even be transformed by the GECC. Complex karyotypes are GECCs, and they are often seen in IAML cells. This may prove that IAML cells can survive from a GECC. Similarly, some *MLL* rearrangements and t(1;22) may be also GECCs, because they are mostly seen in IAMLs. Taken together, we hypothesize that IAML may develop as a consequence of defective development of HSCs/MCs by a genetic disorder.

JMML may develop also as a result of rapid cell transformation of a poor-developed MC. The cell of origin of JMML may be a granulocyte-monocyte progenitor (GMP). Some forms of abnormalities on chromosome 7, such as (+7) and (-7), may be the GECCs that are responsible for the "one-step" transformation of a GMP in JMML. The cell transformation in JMML may be at intermediate-grade, because the leukemia cells in JMML are relatively mature and they can enter bloodstream and infuse into organs.

### XII. Summary

We have discussed in this paper the causing factors and the mechanism of cell transformation of a MC in development of myeloid leukemia. Here we give an outline of our work on studying AML and CML (Figure 5).

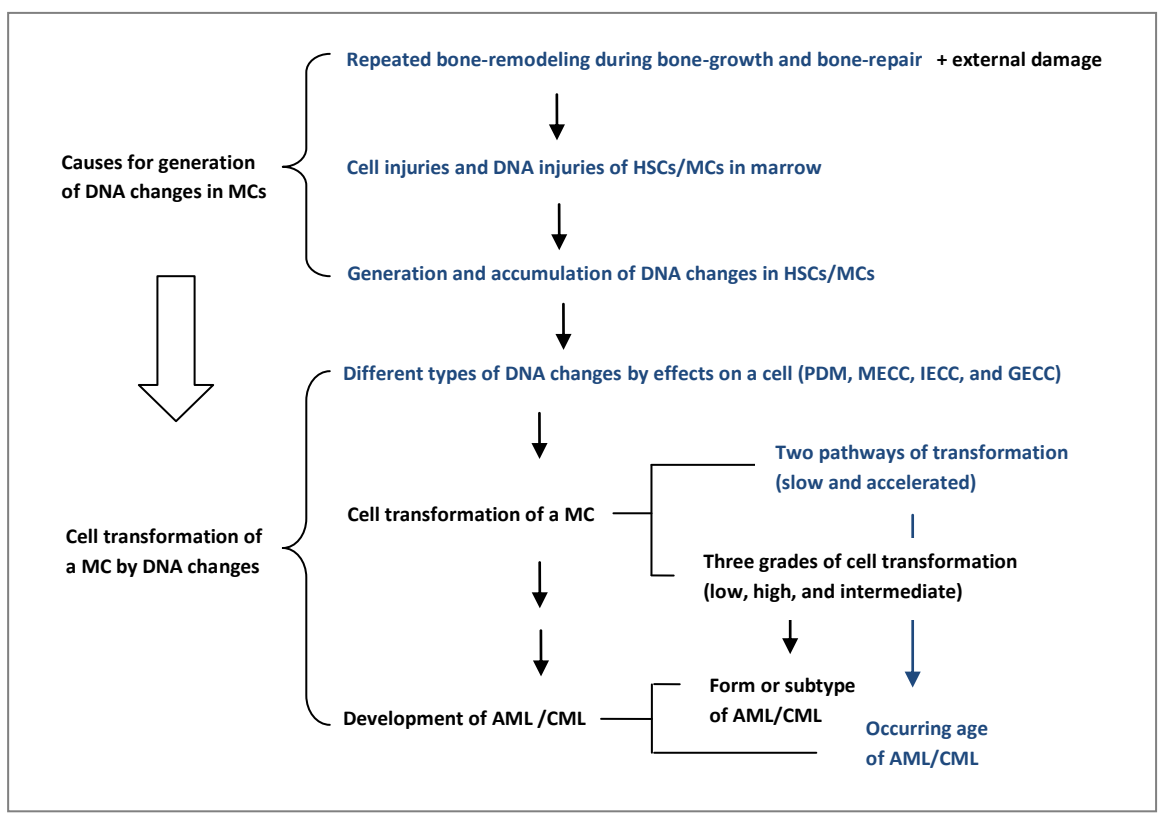

**Figure 5. Our understanding of the pathogenesis of myeloid leukemia**

This is an outline of our work on studying the pathogenesis of myeloid leukemia. DNA changes are triggers for cell transformation. DNA changes are generated in cells as consequences of cell injuries and DNA injuries. In our view, repeated bone-remodeling during bone-growth is a potential source of cell injuries of HSCs/MCs in marrow cavity. Generation and accumulation of DNA changes in HSCs/MCs are results of repeated cell injuries and repeated cell proliferation.

DNA changes have four types by their effects on a cell: PDM, MECC, IECC, and GECC. Different types of DNA changes may drive cell transformations of a MC at different grades (low, high, or intermediate) and via different pathways (slow or accelerated). The form or subtype of AML/CML is determined by the grade of cell transformation of a MC. The age of occurrence of AML/CML is determined by the transforming pathway of a MC. The underlined parts in blue color in the figure are our work.

## XIII. Conclusions

We have discussed in this paper the causes and the mechanism of transformation of a MC. We raise two hypotheses based on our discussion. Hypothesis **A** is: repeated bone-remodeling during bone-growth and bone-repair may be a source of cell injuries of marrow cells including HSCs, MCs, and LCs. Hypothesis **B** is: a MC may have two pathways on transformation: a slow and an accelerated. A transformation via slow pathway occurs at old age; whereas that via accelerated pathway occurs at any age. Therefore, CML and pediatric AML may develop as a result of transformation of a developing MC via accelerated pathway.

MDS and adult AML may develop as a result of transformation of a developing MC via slow pathway. Thus, distinguishing between the two pathways of transformation of a MC can help us to understand why AML, CML, and MDS occur at different ages.

To verify our hypothesis that bone-remodeling is related to the developments of AML and CML, experimental researches can be undertaken to study the association of repeated bone injuries with the incidence of myeloid leukemia in animal models (such as rabbits). Occurrence of leukemia in a child is a tragedy. Our discussion can make us better understand this tragedy. Pediatric AML/CML may develop as a co-effect of three factors: **A.** higher survivability of a MC from DNA changes than a tissue cell; **B.** requiring obtaining fewer cancerous properties of a MC for cell transformation than a tissue cell, and **C.** higher risk of cell injuries of HSCs/ MCs in a child by repeated bone-remodeling for bone-growth. Thus, for reducing the risk of occurrence of AML/CML, avoiding violent sports and bone-injuries may be helpful. Cryopreservation of the HSCs in umbilical cord at birth is advised for the entire population.